\documentclass[12pt,a4paper]{article}

\usepackage{a4wide}
\usepackage[dvips]{hyperref}

\usepackage[dvips]{graphicx}
\usepackage{amsmath}
\usepackage{amssymb}
\usepackage{natbib}
\bibpunct{[}{]}{;}{a}{}{,}
 \newcommand{\um}[1]{\ensuremath{\,\mathrm{#1}}}    
 \newcommand{\particle}[1]{\ensuremath{\mathrm{#1}}}
 \newcommand{\antipart}[1]{\ensuremath{\overline{\mathrm{#1}}}}
 \newcommand{\op}[1]{\ensuremath{\mathsf{#1}}}    
 \newcommand{\vet}[1]{\ensuremath{\mathbf{#1}}}   
 \newcommand{\de}{\ensuremath{\mathrm{d}}}        

\title{\textsf{\textbf{%
Searches for Cosmic Antimatter
}}}

\author{Diego Casadei,\\                 \footnotesize
INFN, Sezione di Bologna,\\              \footnotesize
Via Irnerio 46, I-40126 Bologna, Italy\\ \footnotesize
\url{Diego.Casadei@bo.infn.it}
}

\date{\footnotesize Rev. 2.1 --- \today}

\begin{document}

\maketitle

\begin{abstract}
  We know from experimental high energy physics that whenever matter is
  created, an equal amount of antimatter is also created.  However, we live in
  a large region of the universe where the antimatter can not constitute more
  than a very small fraction of the total mass.  The cosmic antimatter problem
  has been addressed since the beginning of modern cosmology, but no definite
  answer has been formulated despite of the several approaches that can be
  found in the literature.  In this chapter we will make a historical review
  and we will focus on the experimental techniques that has been proposed to
  reveal directly and indirectly the presence of cosmic antimatter in the
  universe.  Indirect searches can be carried on with the measurements of the
  electromagnetic radiation in the microwave and gamma-ray range, and of the
  neutrino flavour, whereas direct searches lay on the measurement of the
  cosmic rays and probe shorter distances.  Finally, the current limits on the
  cosmic antimatter to matter ratio are compared to the sensitivity of future
  experiments.
\end{abstract}

\section{Introduction}

The discovery of ``antiparticle'' solutions of the Dirac's equation in 1929
was followed in 1932 by the experimental discovery of the positron by Blackett
and Occhialini.  With the developement of the accelerator physics, it was
experimentally established that whenever we create new particles in
laboratory, they come in two different forms that are well balanced,
generically called ``matter'' and ``antimatter''.

In particular, the creation (and the annihilation) of fermions is governed by
few conservation laws, the baryonic and leptonic numbers conservation being
the most important ones.  These laws say that if we create fundamental
fermions, each with a positive baryonic (or leptonic) number, in the same
reaction other fermions will be created, each with negative baryonic (or
leptonic) number, in order to keep the total baryonic and leptonic numbers
constant\footnote{In the Standard Model of fundamental interactions, the three
  lepton numbers associated to the electron, muon and tau leptonic doublets
  are separately conserved.}.  In the simplest case, this means that we cannot
create a single fermion: we must create a couple of particle and antiparticle
at least.

When we think about the creation of the present universe, our first attempt
would lead naturally to a symmetric cosmology in which matter and antimatter
are present in the same amount, but astrophysical measurements say that we
live in a big domain that seems to be completely made of matter.  Of course,
the possibility exists that the universe is symmetric on average, consisting
of a collection of homogeneus domains separated by ``walls'' filled with
radiation only.  An important point is then the estimation of the domain size.
If they are of the scale of galaxies or galaxy clusters, we may detect
antimatter cosmic rays (CR) coming from the nearest domain.  On the other
hand, if some antistar exists in our Galaxy we may even detect antinuclei with
$Z\ge2$.

Nevertheless, antimatter cosmic rays do exist and are detected.  The two
species already measured are antiprotons and positrons: they can be produced
by the CR interactions in the inter-stellar medium (ISM) or in the Earth
atmosphere, and they might have a cosmic origin or even be produced by the
annihilation of exotic particles, not present in the Standard Model. However,
the existing measurements do not provide strong hints for primary origin: they
are fully consistent with the secondary production.  On the other hand, the
detection of one anti-helium nucleus would be a striking evidence for the
existence of anti-stars in our Galaxy, because the probability to produce the
anti-alpha particle among the secondaries is negligible.

\section{Antimatter}

What the word \emph{antimatter} means is not simple, thus we start from its
constituents: the \emph{antiparticles}.  If all the characteristics of an
elementary particle (i.e.~a particle without any internal structure), like its
mass, charge and spin are known, then its associated antiparticle is like its
specular image: it has the same mass, but opposite charge and spin.  The best
example is given by the electron (with negative unit electric charge) and its
antiparticle, the positron (with positive charge), whose annihilation at rest
is responsible of the well known 511 keV emission line.

\subsection{CPT theorem}

In modern physics, particles (and antiparticles) are described by the
Relativistic Quantum Theory of Fields, in which a couple of field operators
are capable of destroying or creating one particle of each kind in every given
state.  One can switch between a particle and the associated antiparticle
using the \emph{charge conjungation} operator \op{C}:
\begin{equation}
   \op{C}: \psi(\vet{r},t) \to \overline{\psi}(\vet{r},t)
\end{equation}
where $\psi(\vet{r},t)$ represents a particle quantum field and
$\overline{\psi}(\vet{r},t)$ is the corresponding antiparticle field.

There are two other important discrete operators: the \emph{time reversal}
operator \op{T} and the \emph{parity} operator \op{P} (the spatial inversion):
\[
\begin{split}
   \op{T}:\; &t \to -t \\
   \op{P}:\; &\vet{r} \to -\vet{r} \; .
\end{split}
\]

Even though in general there is no exact symmetry with respect to these
operators, the \emph{CPT theorem} says that every relativistically covariant
quantum field theory, that admits a minimum energy state and obeys the
principle of microcausality (requiring that independent measurements can
always be done on two spacetime points which are outside each other's light
cone), is invariant under the action of \op{C}, \op{P} and \op{T} together,
without any dependence from the order they are applied.

The strict correspondence between particles and antiparticles is a result of
the \op{CPT} symmetry.  In particular, the fact that their masses are exactly
equal is due to the commutative property between \op{CPT} and the Hamiltonian
operator.  In addition the \op{CPT} composite operator is antiunitary: ``it
relates the S-matrix\footnote{In Quantum Field Theory, the scattering process
  is modeled as a very short and intense interaction, that is able to change
  the particle state from the initial free motion to a generally different
  final state (again a free wave).  The S-matrix contains the probability
  amplitude of the transition from any initial state to all the final states,
  thus representing the most complete description of the scattering process
  itself.}  for an arbitrary process to the S-matrix of the \emph{inverse}
process with all spin three-components reversed and particles replaced with
antiparticles'' (quoted from \cite{weinberg95}, p.~183).  This means that the
following two probability amplitudes are equal (an overline denoting
antiparticles):
\begin{equation}
\begin{split}
   A(a_1+a_2+\ldots &\to b_1+b_2+\ldots) = \\ 
                    &A(\overline{b}_1+\overline{b}_2+\ldots \to
                       \overline{a}_1+\overline{a}_2+\ldots)
\end{split}
\end{equation}
(the demonstration of this theorem can be found in the books by \cite{zuber85}
and \cite{weinberg95}, for example).

\subsection{Anti-systems}

The existence of antiparticles, obtained making \op{C} acting upon particles,
does not guarantee the existence of bound systems made with antiparticles.  In
other words, the presence of the \op{C} symmetry alone does not imply that our
system can simply be replaced by another system with antiparticles in place of
particles: to obtain the anti-system we do need the more complex \op{CPT}
symmetry, that involves also the spatial and temporal reflections, changing
indeed the \emph{dynamics} of the system (not only its composition).

The antimatter is formed by compound systems like anti-atoms, that are made of
a cloud of positrons surrounding a nucleus containing antiprotons and
antineutrons.  Due to the \op{C} invariance of the electromagnetic
interactions, all chemical interactions would be the same as ordinary matter,
allowing for macroscopic antimatter agglomerates.

Howewer, one important point is that the electroweak and strong interactions
appear neither to be symmetric with respect to the \op{C} and \op{P}
operators, nor to the composite operator \op{CP}, as found by Cronin and Fitch
in 1964 (for the references to original works, see \cite{weinberg95} or
\cite{maiani96} and references therein).

The experimental discovery of the antimatter, in the sense of bound systems of
antiparticles, was done in 1965 by A.~Zichichi and his collaborators at CERN
\citep{massam65} and by S.~Ting and his collaborators at Brookhaven
\citep{dorfan65}: these groups independently discovered the simplest
antinucleus, the antideuteron.  In 1996 the simplest antiatom, the
antihydrogen, was obtained and studied at CERN and at FERMILAB (see
\cite{blanford97} et references therein).

\section{Cosmic antimatter}

As far as high energy physics experiments are concerned, matter and antimatter
are created exactly in the same amount: the CPT symmetry holds up to a level
of $10^{-12}$, electronic lepton number violations may occur only below
$10^{-12}$ (whereas the limit on the total lepton number, including all
families, is below $10^{-10}$), and the experimental tests on baryon
conservation imply that violations (if any) must be below $10^{-6}$
\citep{pdb2002}. 

What about the matter we see in the universe?  If we think that the
conservation laws hold true during all the history of the universe, we must
admit that somewhere there should be the antimatter that balances the matter
we are consisting of.  If the conservation laws were valid for the whole
universe life, then it is important to explain the presence of non annihilated
matter.

On the other hand, we could state that the baryonic and leptonic numbers were
not conserved in the past, i.e.~that unknown physical processes may have
happened during the first phases of the universe evolution.  Three conditions
were formulated by \citet{sakharov67}, in order to have a non symmetric
baryogenesis: (1) non conservation of the baryonic number $B$ and (2) non
conservation of \op{C} and \op{CP} during a phase in which (3) the cosmic
evolution was out of thermodynamic equilibrium.

Today we can evaluate the ratio between the cosmic background radiation (CBR)
photons and the nucleons (plus antinucleons) number as $\eta \simeq 6 \times
10^{-10}$ \citep{pdb2002}.  The CBR is the result of the annihilation of
particles and antiparticles (assumed to be in a nearly perfect thermal
equilibrium in the very early phase of the universe) when the temperature
became low enough to break the equilibrium between radiation and fermions.  If
all the matter in the universe is of a kind only, then $\eta$ is the relative
difference between the amount of matter and antimatter at the epoch in which
the radiation uncoupled from the fermions.  On the other hand, if the symmetry
is conserved, $\eta$ is the value of the \emph{local} fluctuation in the
quantity of matter and antimatter in that epoch.

The modern picture of cosmic evolution (see for example the recent review by
\citet{stecker02}) is based on the idea that the universe expansion had a very
fast acceleration, the so-called ``inflation'' \citep{guth81,linde82}, during
which a very small (at the quantum scale) bubble was stretched so much that
today it has the dimensions of the visible universe.  The baryogenesis locally
met the Sakharov's conditions, and inflation has probably enlarged homogeneous
domains made of matter or antimatter up to scales comparable with the size of
the visible universe, as supported by recent simulations \citep{cohen98}.

\subsection{Can total annihilation be avoided?}

The Standard Model of fundamental interactions and the Big Bang cosmological
model, when no inflation is considered, unavoidably lead to the prediction of
exponential drop of any inhomogeneity: the equilibrium between the total
amount of matter and antimatter in the universe should be perfect.  However,
observations show that we live inside a (quite large) homogeneous domain made
of matter.  The non zero value of $\eta$ is a problem\footnote{Well, we do
  need this theoretical problem, to exist.}: if the universe is made of matter
only (plus the radiation) we have to understand the mechanism that produced
such a little asymmetry; if the universe is symmetric, we have to understand
how fluctuations could survive and generate the observable structures.

The first attempt to answer to the latter question was due to
\citet{alfven65}.  He considered an ``ambiplasma'' (a fully ionized plasma
consisting of protons, antiprotons, electrons and positrons) and the possible
formation of homogeneus cells separated by the radiation emitted from
``leidenfrost''\footnote{``Leidenfrost'' is the German term used to indicate
  the process in which the water drops ``bounce'' upon a hot layer without
  touching it, sustained by their own vapor pressure.}  leyers in which all
the annihilations happen. \citet{omnes71a} found that this layer is stable
when the magnetic field is negligible: the annihilations cannot disrupt the
separation walls.  In addition, \citet{stecker72} showed that the
perturbations induced by the annihilations could be compatible with galaxy
formation, which would be triggered by the generated turbolence inside a
matter (or antimatter) domain, whose scale has to be that of a galaxy cluster.

\citet{unno74} applied these results to a specific case, trying to explain
quasars as super massive stars composed of matter and antimatter.  If the
antimatter is a little fraction of the total mass, they showed that it would
constitute a domain surrounded by the matter and separated by a leidenfrost
layer, like a bubble inside a star.

To explain how the leidenfrost layer can be able to keep separate matter and
antimatter, \citet{aly78} computed the annihilation rate at the boudary in 3
cases: radiative and plasma eras of Big Bang model, strong magnetic field, two
intergalactic hot domains.  In the meanwhile, \citet{lehnert77,lehnert78}
showed that at interstellar densities no well defined boundary can form in
presence of neutral gas or in a unmagnetized fully ionized plasma.  Only a
magnetized ambiplasma could produce a leidenfrost layer, whose thickness is
proportional to $B T^{1/2}$: for $10^5 < T < 10^8$ K and $B = 10^{-8}$ T, the
wall thickness is about $10^7$~m.

Thus the walls are very thin (of the order of magnitude of the Earth diameter)
and pratically invisible for distant observers.  In analogy with the
geomagnetic field structure, that has well defined zones separated by current
and neutral sheets that can be detected only by spacecrafts traversing them,
\citet{alfven79} pointed out that the whole space is likely subdivided in
cells.

A recent analysis from \citet{dolgov01} emphasized that the behavior of the
domain walls could be different from the leidenfrost process, or more
precisely it could be the opposite: instead of repulsion, the layer could
attract matter and antimatter towards the annihilation region.  Actually,
because electrons, positrons and neutrinos have larger mean free paths than
the domain walls thickness, the energy and pressure density of the
annihilation region decreases, increasing the diffusion of matter and
antimatter towards each other and amplifying the efficiency of the
annihilation.

However, in the framework of the inflation (for a review, see for
example \citet{liddle01}) the problem may even be ill-posed, because
the size of any conceivable antimatter domain would have been enlarged
so much that the domain boundaries are well beyond the radius of the
visible universe.

\subsection{Antimatter domains}

It seems possible that mechanisms exist that could produce the formation of
separated homogeneous domains during the cosmic evolution, thus allowing the
visible universe to be matter-antimatter globally symmetric.  This is an
important point, because it does not require any conservation law breaking,
but we need an estimation of the domain dimensions.  In addition the
possibility exists to have antimatter confined into condensed bodies like
antistars in our Galaxy.

\subsubsection{Uniform domains}

\citet{steigman76} considered homogeneous and uniform domains filled with
matter or antimatter and concluded that they should be at least of the scale
of galaxy clusters, due to the constraints coming from the measured gamma-ray
flux.  The diffuse gamma-ray background flux with energy $E > 100$ MeV, of the
order of $10^{-5}\um{cm^{-1}} \um{sr^{-1}} \um{s^{-1}}$, was used by Steigman
to infer upper limits for the antimatter fraction in the Local Cluster.  For
the hot H II intergalactic medium, his upper limit is $\sim 10^{-7}$, while
inside our Galaxy the limits are more stringent: $10^{-15}$ for galactic
clouds and inter-cloud medium, $10^{-10}$ in the halo.

Steigman estimated that the minimum distance to antimatter domains has to be
10 Mpc, but \citet{cohen98} imposed a much stronger limit, under the
assumption of a baryo-symmetric universe, concluding that the antimatter
domains have to be very distant from us, at least few Gpc (hence they could be
outside our visible universe).  The key point is the smoothness of the cosmic
microwave background (CMB) radiation, that requires density fluctuations below
$10^{-4}$ at scales larger than 15 Mpc, thus implying that, if existing,
matter and antimatter domains must be in close contact.  But the annihilations
products will carry away very efficiently the energy from the contact region,
because electrons, positrons and neutrinos have larger mean free paths than
the domain walls thickness.  Hence the energy and pressure density of the
annihilation region decreases, increasing the diffusion of matter and
antimatter towards each other and amplifying the efficiency of the
annihilation (in contrast with the leidenfrost process) \citep{dolgov01}.

Had this processes happened after the hydrogen recombination, the domain walls
regions would have been strong gamma-ray sources, producing detectable
non-uniformities in the extragalactic gamma-ray spectrum \citep{gao90}.  Non
observation of this background means that any antimatter region must be near
or beyond the horizon.  If the annihilation took place before the hydrogen
recombination, there would be some effect on the CMB energy spectrum, which
must be below the sensitivity of present data (\S\ref{cmb}).

\citet{dolgov01} reviewed few different models, among wich it appears that a
viable model that could account for the existence of domains smaller than the
visible universe is based on ``isocurvature fluctuations'': the initial
baryon/antibaryon asymmetry was zero and started to rise only relatively late,
due to fluctuations in the baryons density.  Baryon (antibaryon) rich regions
cooled faster and diffusing photons from hotter regions had the effect to drag
those regions away, thus providing a way to get separated matter and
antimatter domains.  In this model the annihilations could be weak enough to
create a universe consisting of possibly large domains separated by thin
baryon and antibaryon voids.  Again, how large are those domains?

\subsubsection{Non-uniform domains}\label{small-dom}

The upper limits on the fraction of cosmic antimatter are not so stringent, if
the possibility that it is localized into small domains is considered
\citep{chechetkin82a,chechetkin82b}.  \citet{khlopov98} showed that the
minimum size for such domains is that of a globular cluster (about 1 kpc):
smaller regions have been annihilated during the cosmic evolution.  The mass
of these domains would have probably been condensed into anti-stars during the
galaxy formation, hence the best place to look for an antimatter globular
cluster is the bulge of our Galaxy, where old first-generation stars are
found.

A cosmological model which is able to produce small ($\sim 1$ kpc) antimatter
domains was proposed by \citet{khlopov00b}, who considered a rather narrow
time slice for baryogenesis.  Their mechanism of spontaneous baryogenesis
implies the existence of a complex scalar field carrying baryonic charge with
explicitly broken $U(1)$ symmetry, coexisting with the inflaton field.  During
the inflation, the phase of the complex field behaves as a Nambu-Goldstone
boson, and quantum fluctuations induce different phase tilts in different
regions of the universe.  When the energy is sufficiently low, the field
``rolls down'' the potential and start oscillating, creating baryons and
antibaryons with a proportion which depends on the phase tilt.  If the
baryogenesis happen at the beginning of the inflation, one unavoidable ends
with uniform domains which are enlarged so much that they are today larger
than the visible universe.  However, if it takes place (not much) later, the
final scale of antimatter domains can be at the same time larger than the
minimum ($\sim 1$ kpc) and smaller enough not to produce any signature in the
CMB spectrum.  The result of this model is a non balanced amount of matter and
antimatter in the visible universe, in which the small antimatter domains
contain a very small fraction of the total mass.

\subsubsection{Condensed bodies}

The diffuse cosmic gamma-rays background cannot put an equally stringent limit
to the amount of condensed antimatter bodies, like anti-stars or
anti-planetoids.  \citet{steigman76} inferred for the number of antistars an
upper limit of $10^7$ in our Galaxy (i.e.~$10^{-4}$ of the total stars
number).  This not so stringent limit arises from the fact that antimatter
confined into compact structures like antistars is well separated from the
matter environment and is able to survive longer than in gas clouds.

An antistar is not expected to be a strong gamma-ray emitter, at least if it
does not cross a galactic cloud neither it impacts on other condensed bodies.
\citet{dudarewicz94} give as lower limit on the distance of the nearest
antistar only about 30 pc, and give a $10^{-3}$ upper limit on the fraction of
antistars in M31.  Howewer, they emphasize that the fraction could be of order
unity at the Hubble radius, having superclusters and anti-superclusters
sufficiently well separated, in order to restore the matter-antimatter global
symmetry, even though they conclude that a perfect symmetry appears very
improbable.

\citet{khlopov98} suggested the possibility that antimatter stars could have
survived since the beginning of galaxy formation: they should be searched for
in the globular clusters.  In fact, condensation of an antimatter domain
cannot form an astronomical object smaller than a globular cluster, and the
formation of isolated antistars in the surrounding matter is impossible, since
the necessary thermal instability would finally favor the total annihilation.
Thus antistars can form in an antimatter domain only, and they must constitute
today a whole antimatter globular cluster at least.

In this case, several antistars in the Galaxy could be found in the roughly
spherical globular clusters halo around the Galactic center, that contains old
first-generation stars.  The number of globular clusters is $\sim 10^2$ and
each cluster contains about $10^6$ old stars.  In addition, the antistars are
well separated from the rest of the Galaxy, and the upper limit of $10^{-7}$
calculated by Steigman under the assumption of a uniform distribution may be
well underestimated.

The upper limit to the total mass of one antimatter globular cluster in our
Galaxy, was estimated to be of the order of $10^5 M_\odot$ by
\citet{belotsky98}, who then computed a maximum fraction of antimatter to
matter in cosmic rays of order of $10^{-6}$.  Because current limits on the
anti-helium to helium ratio are at the level of $10^{-7}$, one would today
estimate as $\sim 10^4 M_\odot$ the total mass of an antimatter domain, in the
hypothesis that it belongs to the Galaxy.

Signatures of an antimatter globular cluster inside our Galaxy would be the
presence of \antipart{^3 He} and \antipart{^4 He} in the cosmic radiation, and
the flux of gamma-rays coming from the annihilation of the emitted antiprotons
with the ISM.  Actually, such annihilation would happen pratically at rest,
and produce neutral pions among secondary particles.  The decay of $\pi^0$
would lead to a bump of the diffuse gamma-ray spectrum at about 70 MeV, which
is compatible with EGRET data \citep{golubkov00}.

\section{Searches for cosmic antimatter signatures}

There are few ways in which cosmic antimatter may show itself.  Indirect ways,
which are now considered, are based on the detection of annihilation radiation
in the gamma-ray regime and in the CMB spectrum, on the measurement of the
helicity of photons and neutrinos emitted during non \op{CP} invariant
processes.  The direct way (section~\ref{direct}) is the detection of
antinuclei among cosmic rays.

\subsection{Collision of matter-antimatter bodies}\label{sec-collision}

\citet{sofia74}  considered the collision between
a star and antimatter ``chunks'' ($m \sim 10^{12}$ kg) and found that
the annihilations due to the stellar wind are not important and that
the annihilation rate is limited by the rate at which the matter is
swept out by the chunk due to the stellar radiation.  Thus the impact
with the star cannot be avoided and the chunk penetrates into the star
for $\sim 10^6$ m before eventually evaporating completely.

The chunk would become a hot and expanding antimatter bubble that
would return to the stellar surface due to buoyancy in $\sim 10^2$ s.
Annihilations produce charged and neutral pions, and they decay to
electrons/positrons of 50--70 MeV and gamma-rays of $\sim70$ MeV
(``prompt'' photons).  The photons produced inside the star suffer
$\sim10$ scatterings prior to escape, degrading to energies of
hundreds of keV.  The inverse Compton scattering of those
\particle{e^+}/\particle{e^-} in the stellar atmosphere ($\sim10^4$ K)
will produce a number of $\sim60$ keV photons with time constant of
$\sim10^3$ s (``delayed'' photons).

The signature of such a collision would then be a precursor burst
emission line at $\sim0.5$ MeV (\particle{e^+}/\particle{e^-}
annihilation) with a $\sim70$ MeV continuum lasting 0.1--1 s, followed
by the main annihilation burst at $\sim 100$ keV (10--100 s) and by
the inverse Compton photons ($E \lesssim 100$ keV, $\tau \sim 10^3$
s).  This signature is not very different from the time evolution of
many gamma-ray bursts (GRB). 

The cross section for the chunk capture with relative velocity $v$ at
infinite by a star with mass $M$ and radius $R$ is $\sigma \sim \pi
(2GMR) / v^2$.  A star like the Sun, for $v \sim 10^4$ m \um{s^{-1}}
has $\sigma \sim 10^{23}\um{m^2}$, and during 1 year sweeps up a
volume $V = \sigma v t \sim 10^{34}\um{m^3}$.  In the Galactic disk, a
sphere of radius 100 pc ($V=10^{56}\um{m^3}$) contains about $10^5$
stars, then there will be a collision every $\sim 10^{15}$ years (much
greater that the Univerge age).

A similar way was followed by \citet{sofia76},
who considered the collision between antimatter asteroids and the Sun,
while \citet{alfven79} considered star-antistar
collisions, possibly ending in ``ambistars'', i.e.~stars with matter
and antimatter whose annihilations contribute with the thermonuclear
fusion processes to the total emitted power.

The collision with small ($r<10$ km) bodies in the Solar system and
the encounter of clouds with antimatter clouds were considered by
\citet{rogers80,rogers82}.  They found that very small antimatter
objects in the Solar system would produce a gamma-ray flux of the
order of $10^{-10}\um{cm^{-2}} \um{s^{-1}}$, too low to be detectable.
In addition, different clouds will not merge.  Instead a thin
``leidenfrost'' layer will form ($\sim 10^9$ m, compared to $\sim
10^{15}$ m scale length for clouds), and annihilation will burn only a
very small ($\sim 10^{-12}$) fraction of the total mass, resulting in
less severe constraints for gamma-rays emission than those considered
by \citet{steigman76} (but this argument may not be valid, as
\citet{dolgov01} pointed out).  \citet{fargion01} considered
antimatter meteorites in the solar system, obtaining a limit of
$10^{-9}$--$10^{-8}$ on the antimatter to matter ratio.

Actually all the interactions between antistars and the matter in our
Galaxy are very weak until they remain in a bound system like a
globular cluster (this is indeed the reason why they could have
survived until now).  Thus it make sense \citep{foschini01} to consider
the possibility that antistars escape from their cluster, wander
through the Galaxy and possibly interact with the galactic matter.

Following \citet{binney87}, a star (or antistar) can escape from a cluster in
two ways: \emph{ejection}, in which the escape speed is gained in a single
close encounter with another star, and \emph{evaporation}, in which several
distant encounters produce a gradual velocity increase.  The former process is
negligible when compared to the latter, whose characteristic time can be
roughly estimated as $t_\text{ev} \approx 100 \, t_\text{re}$, where the mean
relaxation time $t_\text{re} = (3\times10^7)$--$(2\times10^{10})$ years for a
globular cluster.

We can compute the number of stars in a cluster as:
\begin{equation}
   N(t) = N_0 \exp{(- t/t_\text{ev})}
\end{equation}
and the time $t_1$ elapsed before one star can escape is found by solving the
equation $N_0 - N(t) = 1$, that is:
\begin{equation}
   t_1 = - t_\text{ev} \ln [(N_0 - 1)/N_0] \; .
\end{equation}
The number of stars in a globular cluster is $N_0 \sim 10^6$, hence we get
$t_1 = (3\times10^3)$--$(2\times10^6)$ years.

Because this time is much shorter than the age of the Galaxy, it is
very likely that, if at least one of the galactic globular clusters is
made of antimatter, there are many (possibly hundreds) antistars
wandering in the roughly spherical volume in the center of the Galaxy
occupied by the globular clusters.

Those antistars may interact wih a matter cloud, star or smaller
compact body.  An important effect that has to be considered when
gaseous material is accreting into an antistar is that the equilibrium
between the gravitational and radiation pressure is reached at higher
power than the ``Eddington luminosity''
\begin{equation}
   L_\text{Edd} = \frac{4\pi G M m_\text{p}c}{\sigma_\text{T}}
        \approx 1.3\times10^{38} \frac{M}{M_\odot} \; \um{erg}\um{s^{-1}}
\end{equation}
($M_\odot = 2\times 10^{30}$ kg is the solar mass) because when annihilation
photons are considered, the Thomson cross section $\sigma_\text{T}$ has to be
substituted with the relativistic Klein-Nishina formula: the cross section,
for photon energies much higher than the electron rest mass energy
($m_\text{e}c^2 = 511$ keV), can be approximated by
\begin{equation}
   \sigma_\text{KN} \approx \frac{\pi r_\text{e}^2}{\epsilon} \left(
        \ln 2 \epsilon + \frac{1}{2} \right) , \; \epsilon \gg 1
\end{equation}
where $r_\text{e} = 2.82 \times 10^{-15}$ m is the classical electron
radius and $\epsilon = (\hbar \omega )/(m_\text{e}c^2)$ is the ratio
between the photon energy and the electron rest mass energy.  For
50--70 MeV photons, typically produced by the decay chains of the
charged and neutral pions arising from the nucleon/antinucleon
annihilation, the Klein-Nishina formula gives for the cross section
46--61 smaller values than the classical Thomson value
($\sigma_\text{T} = 8\pi r_\text{e}^2/3 = 0.665 \times 10^{-28}
\um{m^2} = 0.665 \um{barn}$).

The annihilation photons would then escape almost freely from the
accretion disk, taking away a considerable fraction of the total
emitted energy: the net effect is that the power emitted by a
matter-accreting antimatter-star can become much greater than the
usual accretion case, especially in the gamma-ray regime, where normal
stars have negligible emission.

On the other hand, the rare star/antistar head-on collisions would
produce an intense energy release for few seconds, due to the surface
annihilations, before merging and reaching a (probably
super-Eddington) stationary luminosity \citep{foschini01}.  These close
encounters may appear as galactic GRB and, in the very rare case of a
complete mixing, they might produce an ``ambistar'' fueled by
antiproton-proton annihilations, a blue supergiant with strong
$\gamma$-ray emission.

Hence, a possible search for matter-antimatter accretion systems could
be carried on by comparing optical, X-ray and $\gamma$-ray
luminosities of Galactic sources: the signature would be an excess of
emitted power in the $\gamma$-ray range.

\subsection{Polarization of electromagnetic emission}\label{sec-pol}

With a completely different approach, \citet{cramer77} stressed out
that in addition to direct annihilation, antistars may be
distinguishable by the polarization properties of their
electromagnetic emission.  In fact the ordinary thermonuclear
reactions which occur in stars systematically convert protons into
neutrons through the weak-interaction process of $\beta^+$ decay and
electron capture.

When positrons ($\beta^+$) are emitted, they are preferentially in a
``right'' elicity state of strength $v/c$.  Their bremsstrahlung
emission is then preferentially right-circularly polarized.  The same
is true also for the forward going annihilation photon.  In antistars,
antiprotons are converted into antineutrons, producing electrons in a
``left'' elicity state.  The photons produced by those electrons are
then preferentially left-circularly polarized.

During normal star processes, the photons loose the initial
polarization state while diffusing out of the star, but during a
supernova explosion the photons produced by the \particle{^{56}Ni}
decay chain could be detectable.  \particle{^{56}Ni} decays by
electron capture to \particle{^{56}Co}, which decays by electron
capture or positron decay to \particle{^{56}Fe}.  The emitted
positrons will radiate through bremsstrahlung polarized photons at the
surface of the ejected material.  These gamma-rays may then escape and
be detectable.  Hence, a measurement of their degree of polarization
could tell us the nature of their origin.

Two effects can be exploited to measure the polarization of a $\gamma$-ray
photon beam: Compton scattering and pair production.  The differential Compton
cross section of a linearly polarized beam is:
\begin{equation}
 \frac{\de \sigma}{\de \Omega} = \frac{r_\mathrm{e}^2 \beta^2}{2}
   \left ( \beta + \frac{1}{\beta} - 2 \sin^2 \theta \cos^2 \phi \right )
\end{equation}
where $\theta$ is the scattering angle from the incident photon direction,
$r_\mathrm{e}$ is the classical electron radius and $\beta$ is the ratio
between the scattered and incident photon energy.  The polarization is
detected by the difference between the number of photons with a given
azimuthal angle $\phi$ and the number of photons scattered in perpendicular
directions.  Using a two-dimensional germanium strip device to measure the
energy and the position of each interaction, \citet{kroeger98} proposed a
small detector (with geometrical acceptance of few tens of \um{cm^2} sr) whose
sensitivity is peaked roughly at 100--200 keV (maximum efficiency of order of
10\%).

The azimuthal dependence of the cross section for electron/positron production
can be written
\begin{equation}
  \sigma (\psi) = \frac{\sigma_0}{2\pi}
   [1 + P R \cos 2(\psi - \psi_0)]
\end{equation}
where $\psi - \psi_0$ is the relative angle between the pair plane and the
incident electric field vector, $\sigma_0$ is the total cross section, $P$ is
the fractional polarization, and $R \sim 1$ is a numerical factor expressing
the inherent asymmetry of the process. This process is useful for polarization
measurements above few tens of MeV, if the detector can measure the initial
part of the track of two leptons: the pair produced by the incident photon
must be able to propagate without being affected by the Coulomb scattering in
order to find the plane defined by electron and positron tracks with good
angular resolution.  A possibility is to use a low density gas filling a large
volume and position sensitive pixels with very small pitch ($\sim 100 \,\mu$m)
\citep{bloser03}.

\subsection{Effects on the cosmic microwave background radiation}\label{sec-cmb}

The most favored baryogenesis scenario today is based on a very non-uniform
distribution of matter and antimatter, the latter forming very small domains
(see \S\ref{small-dom}).  To obtain small islands of antimatter, the
baryogenesis should have been happened in a small time window during
inflation: a very early process would have produced too large domains, whereas
they have disappeared if produced too late.

The minimum scale $R_{\mathrm{c}} \simeq (10^{-5}$--$10^{-4}) \, \zeta^{-1}
(z/z_{\mathrm{rec}})^{1/2} \, r_{\mathrm{h}}(z_{\mathrm{rec}})$, where $\zeta$
is the ratio between the anti-baryon density inside the region and the baryon
density outside, and $r_{\mathrm{h}}(z_{\mathrm{rec}})$ is the horizon scale
at the recombination \citep{naselsky04}.  At the recombination era
$t_{\mathrm{rec}} \simeq \frac{2}{3} (\Omega_{\mathrm{m}} H_0^2)^{-1/2} \,
z_{\mathrm{rec}}^{3/2}$, where $z_{\mathrm{rec}} \sim 10^3$, $H_0 = 100 h = 73
\pm 3$ km \um{s^{-1}} \um{Mpc^{-1}} is the present value of the Hubble
parameter and $\Omega_{\mathrm{m}} = 0.127^{+0.007}_{-0.013}$ is the fraction
of the universe mass density due to baryons and dark matter.  At this time,
the total baryonic mass is of the order of $10^{19} M_\odot$, and the
antimatter domains could have masses of the order of $(10^4$--$10^7) M_\odot$.

On the other hand, had antimatter domains occupied a non neligible volume
fraction at the nucleosynthesis epoch (at $t \simeq$ 300-1000 s), the light
elements abundances would have been different from present values (which
satisfy all predictions of the Standard Big Bang Nucleosynthesis model).

Matter-antimatter annihilation right before and during hydrogen recombination
would distort the CMB spectrum in different ways, depending on the epoch and
on the spatial distribution of the antimatter domains.  In particular, a non
uniform distribution would have induced structures in the CMB polarization
map.  The physical processes relating the annihilation to the CMB photons are
bremsstrahlung and inverse Compton scattering of e$^+$/e$^-$ produced by the
annihilation, and electromagnetic cascades initiated by the high energy
annihilation charged products.

\subsection{Supernovae neutrinos}\label{sec-neutrino}

Finally, \citet{barnes87}  suggested that the
initial neutrino bursts from a supernova could reveal whether the
source is made of matter or antimatter.  In the first 2--10 ms the
neutronization reaction $\particle{e^-} + \particle{p} \to
\nu_\text{e} + n$ produces a $\sim 10^{52}$ erg burst of $\sim10$ MeV
neutrinos, whose flux cuts off abruptly when the infalling matter
achieves sufficient density to trap them.  This dense infalling matter
comes to thermal equilibrium, in which all neutrino flavours are
produced.  Neutrinos and antineutrinos, approximatively in the same
number, carry away 99\% of the binding energy of the newly formed
neutron star.

The electron neutrinos (and antineutrinos) suffer more scatterings
than muon and tau neutrinos, and escape with a mean energy of $\approx
10$ MeV, roughly half than the muon and tau neutrinos mean energy.  On
the other hand, the produced $\nu_\text{e}$ (and
$\antipart{\nu}_\text{e}$) number is roughly twice the $\nu_\mu$ or
$\nu_\tau$ numbers.  The net effect is that the energy of thermal
neutrinos is equally divided among the three flavours.

In water \v{C}erenkov detectors, like Kamiokande, SuperK and IMB, all
neutrino flavours may interact by $\nu/\text{e}$ scattering, while
electron antineutrinos have an additional channel, the inverse
$\beta$-decay on the hydrogen nuclei (the interaction cross section
for oxygen is negligible): $\overline{\nu}_\text{e} + \text{p} \to
\particle{e^+} + \text{n}$.

The ratio between the $\nu_\text{e}$ emitted during the burst phase
and the number expected from the thermal phase is $r=0.01$--0.03 and
the expected counting rate for the 10 MeV electron neutrinos and the
20 MeV muon and tau neutrinos follows the proportion:
\begin{equation}\label{eq_nu}
  \overline{\nu}_\text{e} \text{p} :
  (\text{all thermal } \nu, \overline{\nu}) \text{e} :
  (\text{burst } \nu_\text{e}) \text{e} =
  10 : 1.1 : 3.3 r  \; .
\end{equation}

If the progenitor star is made of antimatter, an important difference
arises with this picture: the initial burst is due to the
antineutronization reaction $\particle{e^+} + \antipart{p} \to
\antipart{n} + \overline{\nu}_\text{e}$ and the burst contains
electron antineutrinos rather than neutrinos.  The
$\overline{\nu}_\text{e}$ cross section in water is 18 times higher
than the $\nu_\text{e}$ one, and the proportion (\ref{eq_nu}) has to
be replaced with:
\begin{equation}
  \overline{\nu}_\text{e} \text{p} :
  (\text{all thermal } \nu, \overline{\nu}) \text{e} :
  (\text{burst } \overline{\nu}_\text{e}) \text{p} =
  10 : 1.1 : 60 r  \; .
\end{equation}

Thus (6--20)\% of all oberved events from an antimatter supernova are
expected to occur within the first few milliseconds.  In addition, the
$(\overline{\nu}_\text{e} \text{p})$ reaction produces electrons with
nearly isotropic cross section, while the elastic scattering
$(\nu_\text{e} \text{e})$ is peaked forward.  Hence the expected
signature for an antimatter source is an initial burst in a water
\v{C}erenkov detector with isotropic distribution.

From supernova SN 1987A, located in the Large Magellanic Cloud at
$\sim 55$ kpc from Earth, 11 and 8 events were registered by
Kamiokande II and IMB respectively.  Due the too low statistics, it is
impossible to distinguish between the expected $\sim 2$
$(\overline{\nu}_\text{e} \text{p})$ events in case of antimatter star
and the expected $\sim 0.1$ $(\nu_\text{e} \text{e})$ events
corresponding to a matter progenitor star.  The first event
registered by Kamiokande is forward peaked and if it is attributed to
the burst it may prove that SN 1987A was produced by a matter
progenitor star (see \cite{barnes87} and references therein).

\section{Direct detection}\label{direct}

As we have seen, indirect evidences of cosmic antimatter might be found in the
measurements of neutral particles (photons and neutrinos): low-energy photons
of the CMB would carry on information about the antimatter distribution in the
early phase of cosmic evolution (\S\ref{sec-cmb}), whereas high-energy photons
might reveal the presence of antimatter in the modern era.  In particular,
ambistars (\S\ref{sec-collision}) would appear as normal stars with excess of
emission power in the gamma-ray regime, whose formation could be preceded by a
gamma-ray burst, whereas antistar explosions would emit high-energy photons
with a different polarization state (\S\ref{sec-pol}) and a large number of
antineutrinos (\S\ref{sec-neutrino}).

However, direct detection of CR antihelium nuclei would be the most compelling
indication of the existence of cosmic antimatter: \antipart{He} could be of
primordial origin or even be produced by the antiproton fusion in the core of
an antistar.  Antihydrogen is of course expected as the most abundant element
of antimatter domains, but secondary \antipart{p} production in CR
interactions with ISM is an overwhelming source of background for any
conceivable cosmic antimatter search.  The measurements of positrons are even
less significative for this search, because positrons (and electrons) are
commonly produced during the CR propagation in the ISM, and in addition they
loose energy very rapidly, making impossible to probe distances of
cosmological interest.

\citet{khlopov98} suggested the possibility that antimatter globular
clusters could have survived since the beginning of galaxies
formation.  The idea that one antimatter globular cluster may be
present in our Galaxy refreshed the interest into the possible
observation of cosmic antimatter effects.

There are several possible ways in which such an antimatter globular cluster
could manifest itself: its e.m.~emission may show anomalous circular
polarization at all wavelengths, unrelated to any linear polarization which
may be present (\S\ref{sec-pol}); their antistar wind would hardly produce
detectable reactions with the galactic ISM but they may interact with matter
clouds, stars or smaller bodies (\S\ref{sec-collision}).  But the most
important effect may be the detection of antinuclei with $Z>2$, that were
produced only in negligible quantities during the primordial nucleosynthesis.


If a non zero amount of antimatter did survive the primordial annihilation, it
is reasonable to expect that its composition will be similar to that of
ordinary matter.  Hence, we may think about cosmic antimatter domains as
composed by protons, positrons, antihelium nuclei, few isotopes of
antihydrogen and antihelium, with negligible quantities of heavier antinuclei.

Antistars could have formed inside antimatter domains exactly in the same way
as ordinary stars formed in matter domains.  Thus we may expect that nuclear
reactions happen inside antistars, similar (apart from photon polarization and
antineutrino production, as seen in \S\ref{sec-pol} and \S\ref{sec-neutrino})
to ``normal'' reactions: proton-proton and C-N-O chains.

As for matter domains which contains stars and galaxies, antimatter nuclei
would be injected in the ISM, where they would be accelerated as cosmic rays,
and a fraction of them (that depends on particle momentum and distance from
us) could escape from those domains and reach our Galaxy, where they would
continue to diffuse for a long time before annihilation can happen, because
the interaction length ($\sim 60 \um{g/cm^2}$ for protons and antiprotons) is
greater than the escape length ($\sim 5 \um{g/cm^2}$).  If anti-stars exist in
the Galaxy, the probability to detect antimatter cosmic rays is of course
larger.

Thus, a finite probability exists that cosmic ray detectors in the Solar
system may reveal cosmic antimatter.  Actually, such instruments would
certainly detect antiparticles produced by the interactions of cosmic rays
with the interstellar medium.  This background can be completely overwhelming
for certain kinds of cosmic antiparticles, but this is not the case for
antihelium and heavier antinuclei.

\subsection{Positrons and antiprotons}

Protons are the most abundant particles among cosmic rays, and CR
electrons are about 1\% of protons.  Very likely their antiparticles
would be the most abundant species in antimatter domains, and we may
expect that they would constitute the greatest antimatter fraction
among cosmic rays detected on the Earth.

Other sources of antiprotons and positrons are the reactions of cosmic rays
with the interstellar medium.  In fact, among the secondary particles produced
by energetic inelastic scatterings between two protons (the most abundant
species both in CR and ISM) or a proton and a nucleus, the most abundant ones
are mesons, like pions and kaons, and antiprotons.  In addition, while neutral
pions dacay into energetic photons ($E_\gamma = 70 \um{MeV}$ in their center
of mass system), charged pions decay into muons and electron-positron pairs
(also produced by muon decays), so that the secondary production of
antiprotons and positrons is a quite common process.

Like electrons, positrons have short radiation length and suffer heavy
energy losses during propagation in the ISM, hence there is no
possibility that CR positrons be of cosmic origin: they are produced
by the interactions of cosmic rays with the interstellar medium.

\begin{figure}[t!]
 \centering
 \includegraphics[width=0.7\textwidth]{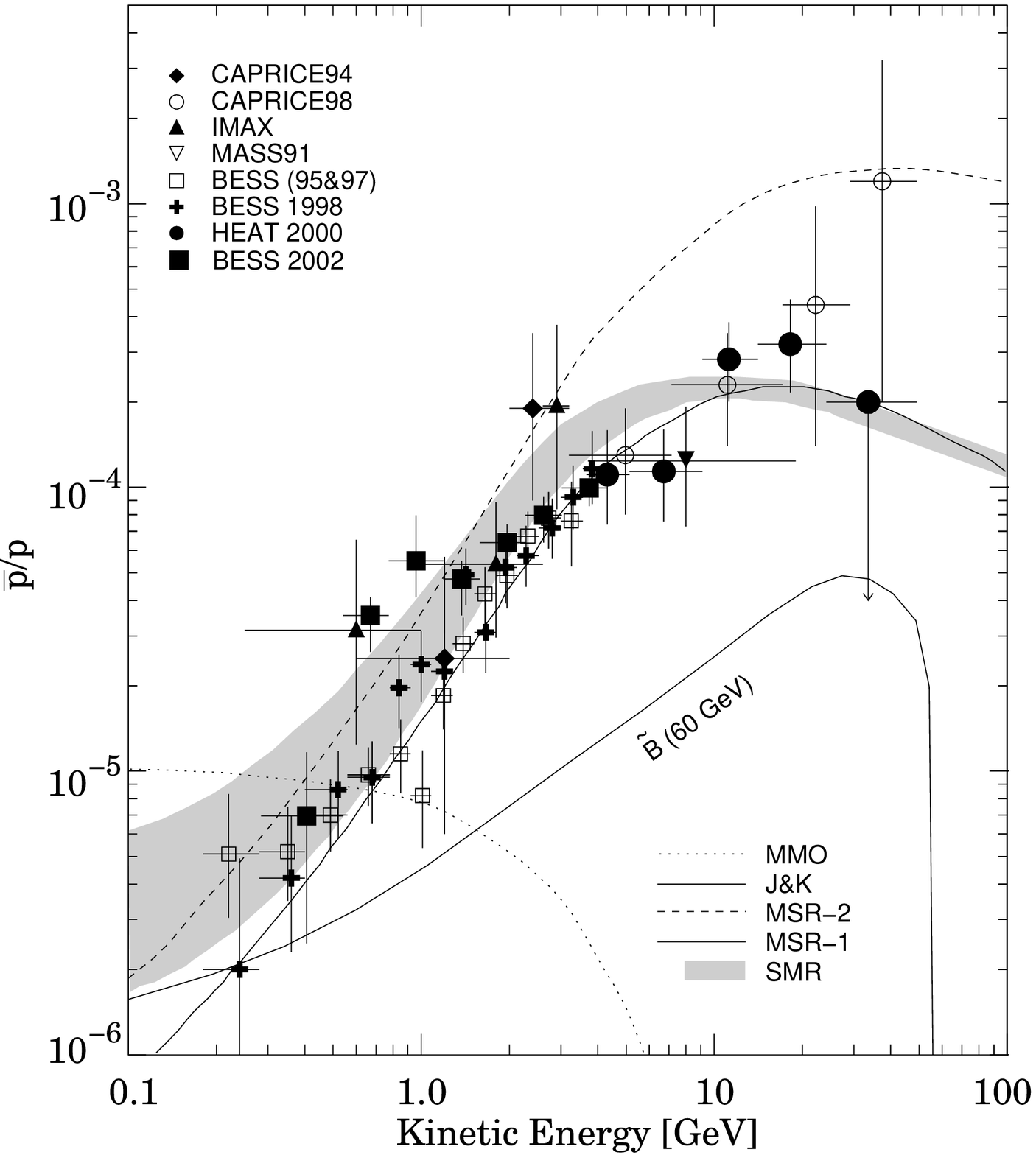}
 \caption{Experimental results on the CR antiproton to proton ratio
    \citep{beach01,haino05}.}\label{antiprot}
\end{figure}

On the other hand, antiproton production is hardly disfavoured for
energies below 2 GeV for kinematical reasons, so that the secondary
antiproton spectrum should have a characteristic peak around 2 GeV
(for higher energies, it is the primary proton spectrum that goes down
as $\sim E^{-2.8}$, while the \antipart{p} production yield is almost
constant).  Thus, cosmic antiprotons (and antiprotons from exotic
sources as dark matter particle annihilation \citep{bergstrom99}) may
be searched at low energies or above the secondary peak.
Figure~\ref{antiprot} shows the experimental results.

\subsection{Antihelium and antinuclei}

Helium is the second most abundant species in the universe: about 25\% of the
total baryonic mass, and about 20\% of CR particles are \particle{He} nuclei.
\particle{^4He} nuclei can be of cosmic origin (produced during primordial
nucleosynthesis) or of stellar origin (produced by the proton-proton and C-N-O
nuclear chains).  After their acceleration, helium nuclei propagate through
the Galaxy for a time similar to the proton propagation time (about $2 \times
10^7$ years), and may interact with the interstellar medium, producing
\particle{^3He} isotopes by spallation.

Similarly, we expect that the greatest fraction of CR antinuclei
(after antiprotons) is constituted by antihelium isotopes.  Actually,
the possible detection of \antipart{He} would be a striking
demonstration that antimatter plays a cosmic role, as annihilation
remnants wandering through the Galaxy or in form of antistars: the
secondary production probability of $^3$\antipart{He} by cosmic ray
interactions with the ISM was estimated to be of order $10^{-13}$
\citep{chardonnet97} and the probability for secondary $^4$\antipart{He}
is much lower.

While antihelium may be of cosmic or (anti-)stellar origin, the
detection of antinuclei could be explained only as a demonstration
that antistars do exist in our Galaxy (or in some nearby galaxy).
Among the possible isotopes, the best candidates for this antimatter
search are $^{12}$\antipart{C}, $^{14}$\antipart{N} and
$^{16}$\antipart{O}, because they are the most probable production
results (after $^4$\antipart{He}) of nuclear reactions fueling
antistars.

\begin{figure}[t!]
\centering
\includegraphics[width=0.7\textwidth]{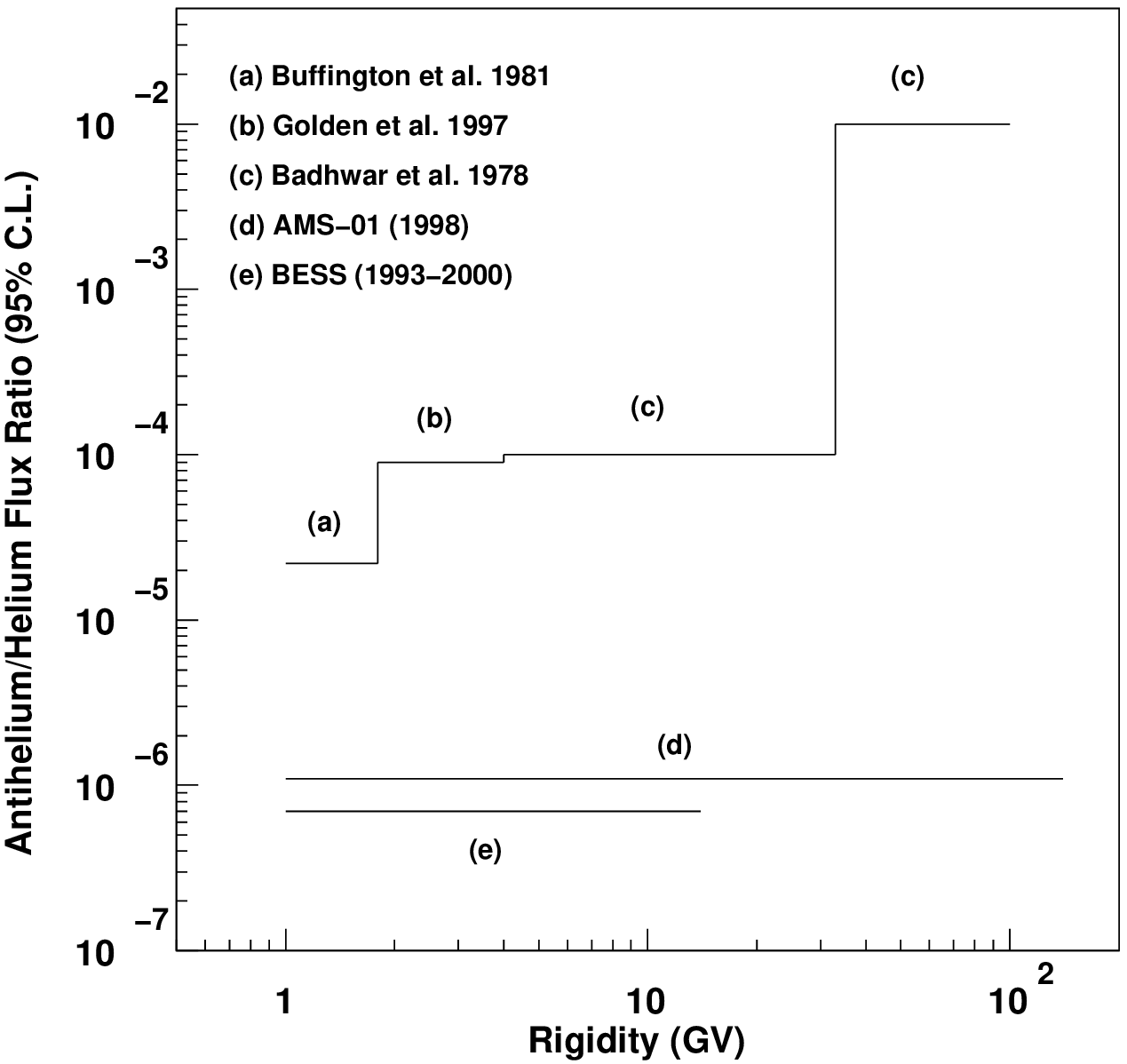}
\caption{Experimental results on the CR antihelium-to-helium flux
ratio \citep{buffington81,golden97,badhwar78,ams99,sasaki01}.}
\label{antiHe-ratio}
\end{figure}

\begin{figure}[t]
\centering
\includegraphics[width=0.7\textwidth]{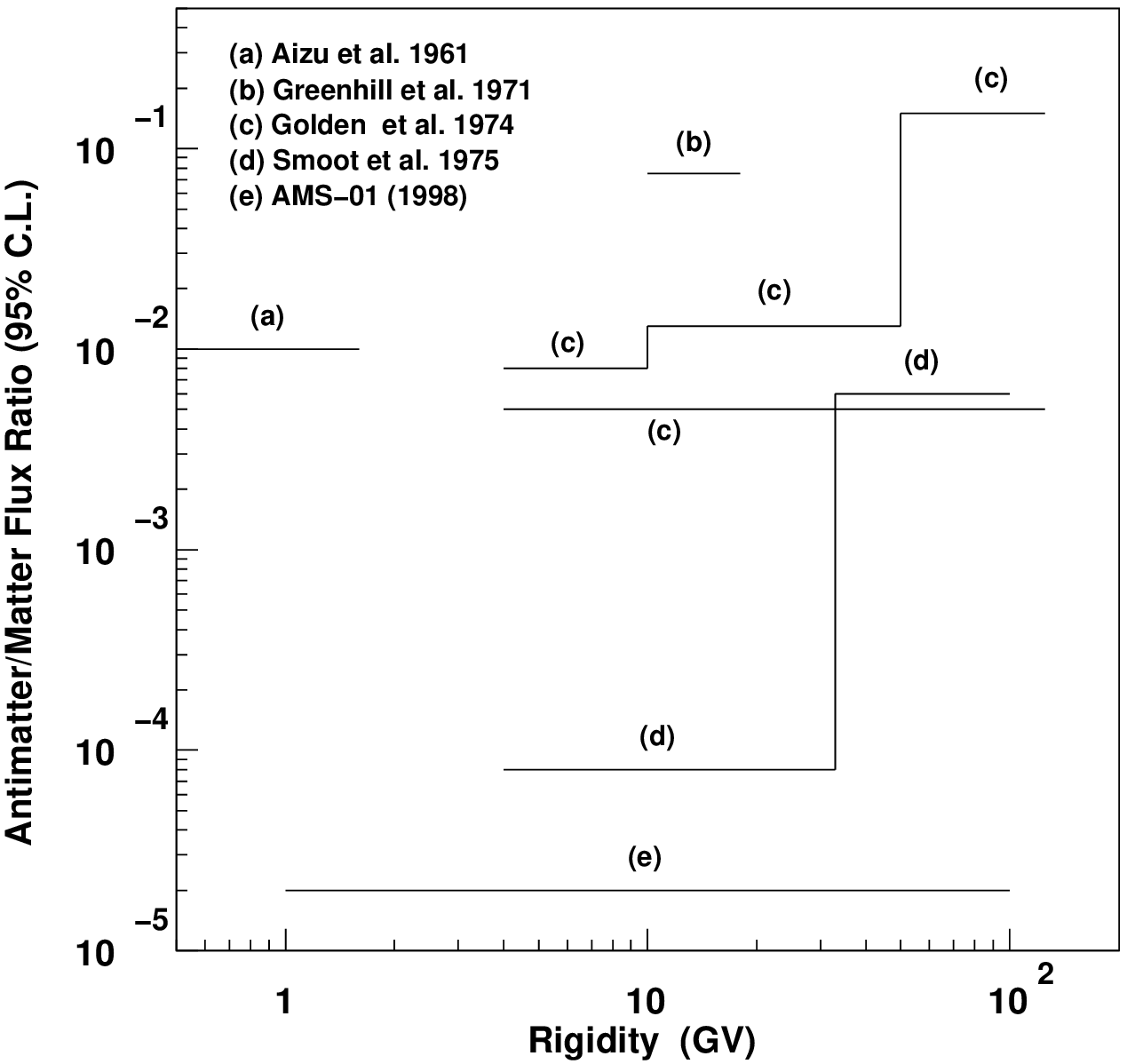}
\caption{Experimental results on the CR antimatter-to-matter flux
ratio for atomic numbers $Z>1$
\citep{aizu61,greenhill71,golden74,smoot75,cristinziani01}.}
\label{antim-ratio}
\end{figure}

Figures \ref{antiHe-ratio} and \ref{antim-ratio} show the experimental upper
limits found by balloon and space experiments on the cosmic ray anti-helium to
helium flux ratio and antimatter (i.e.\ antinuclei) to matter flux ratio,
respectively.

\section{Summary and perspectives}

The problem of the possible existence of primordial antimatter is
still open, despite of the several attempts that has been tried out in
the past 40 years.  From the theoretical point of view, the
alternatives are:
\begin{enumerate}
 \item symmetric cosmology with matter/antimatter fluctuations either
 on small scales, producing homogeneous galaxy clusters, or on very
 small scales, that could lead to the formation of clusters of
 antistars; 

 \item symmetric cosmology with matter/antimatter fluctuations on
 scales that became larger than the size of galaxy clusters, and may
 be even larger than the visible universe;

 \item asymmetric scenario involving possibly new and unknown aspects
 of the fundamental interactions, that were important during the early
 stages of the cosmic evolution and produced a $\sim 10^{-10}$ excess
 of particles over antiparticles.
\end{enumerate}

From the experimental point of view, in the first case we have chances to find
evidence of the existence of cosmic antimatter either indirectly, looking at
the gamma-ray measurements, or directly, using accurate cosmic ray
spectroscopy techniques.  The direct detection of antinuclei is out of the
possibility of cosmic ray experiments in the second scenario, whereas
gamma-ray and neutrino searches can still be carried on.  Of course, in the
last possibility nothing more than lower limits on the size of our homogeneous
domain can be found by astrophysics experiments: only laboratory evidence of
new physics could give hints on asymmetric cosmologies.

The different experimental techniques that will be shortly reviewed are based
on the direct detection of antinuclei among cosmic rays, or on measurements of
the gamma-ray polarization and of the neutrino flavour during supernova
events.

\subsection{Cosmic ray experiments}

The direct detection of antinuclei, that requires a magnetic spectrometer to
distinguish between positive and negative charge, will be possible only below
1 TeV/nucleon with present and approved future experiments: BESS Polar
\citep{bess04} with long duration balloon flights will reach a
\antipart{He}/He sensitivity of less than
$10^{-7}$; PAMELA 
\citep{pamela02} and AMS-02 
\citep{ams02} with longer satellite missions will reach sensitivities of the
order of $10^{-8}$ and $10^{-9}$, respectively.  The discovery of antihelium
nuclei would be a very strong suggestion that antistars do exist in our Galaxy
(in this energy range the curvature radius of a charged particle is too small
to have a non negligible escape probability).  Hence direct searches will be
useful only to test the hypothesis of a symmetric cosmology with
matter/antimatter fluctuations on very small scales.

In order to test the hypothesis of fluctuations on larger scales, one
must rely on measurements of neutral particles, that are not deflected
by magnetic fields and follow the shortest path from the source to the
observer.  For example, a possible search for matter-antimatter
accretion systems could be carried on, as explained in
\S\ref{sec-collision}, by comparing the power emitted in the optical,
X-ray and $\gamma$-ray ranges.

\subsection{Gamma-ray polarimetry}

In \S\ref{sec-pol} we saw that the polarization of $\gamma$-ray photons
emitted in supernova explosion is expected to be different for matter and
antimatter progenitors.  In addition, nearly all emission mechanisms can
produce linearly polarized emission, though the polarization angle and the
degree of polarization depend on the source physics and geometry.

Synchrotron radiation, produced by relativistic electrons spiraling around
magnetic field lines, and curvature radiation, produced by lower energy
electrons following curved magnetic field lines, produce linearly polarized
emission whose polarization angle traces the field direction and the degree of
polarization is independent from the energy.  On the other hand, Compton
scattering of $\gamma$-rays on ambient electrons produces radiation whose
polarization degree depends on the energy and scattering angle.  These
processes are expected to dominate the high-energy radiation of gamma-ray
pulsars, gamma-ray bursts, supernova remnants and active galactic nuclei
\citep{bloser03}.

Recently, the first astrophysics measurement of $\gamma$-ray polarization has
been reported by the RHESSI experiment \citep{coburn03}, even though other
authors stated that a re-analysis of the same data shows no polarization at
all \citep{rutledge04}.

Though the IBIS instrument on
INTEGRAL 
\citep{integral04} has some polarization sensitivity, the next
generation experiments
AGILE 
\citep{agile99} and GLAST 
\citep{glast02} will have a negligible sensitivity to the polarization of
gamma-rays \citep{bloser03}.  However, a number of new detectors has been
recently proposed to carry on polarization measurements of astrophysical
sources.  In addition to the two proposed detectors mentioned in
\S\ref{sec-pol}, other examples are:
GRAPE, using low-$Z$ organic scintillators surrounded by high-$Z$ inorgnic
scintillators to detect Compton scattering of 30--300 keV photons
\citep{grape}; POLAR, a bundle of plastic scintillator ``spaghetti'' operating
in the 10-300 keV range \citep{polar}; NeXT/SGD, with CdTe and Si
semiconductor technology, in the 0.1--10 MeV range \citep{next-sgd}.  High
energy photons polarimetry is likely to be one of the most exciting frontiers
in astronomy for the near future.

\subsection{Cosmic microwave background measurements}\label{cmb}
 
CMB measurements made by WMAP during the first year \citep{bennett03} do not
show any evidence for antibaryon contamination, and can only be used to infer
limits on the parameters of cosmic evolution models, as shown by
\citet{naselsky04}.  Results from the three years data set are currently being
pubblished, and suggest that CMB temperature, polarization and small-scale
structures fit into a six-parameters cosmological model, together with
light-element abundances, large-scale structure observations and the supernova
luminosity to distance relationship \citep{spergel06}.  The best values for
the cosmological power-law flat $\Lambda$CDM\footnote{$\Lambda$CDM is a
  cosmological model with cold dark matter and a dominant contribution from
  the ``vacuum constant'' $\Lambda$.} model are: $\Omega_{\mathrm{m}} h^2 =
0.127^{+0.007}_{-0.013}$ (matter and dark matter density),
$\Omega_{\mathrm{b}} h^2 = 0.0223^{+0.0007}_{-0.0009}$ (baryon density), $h =
0.73 \pm 0.03$ km \um{s^{-1}} \um{Mpc^{-1}} (Hubble parameter),
$n_{\mathrm{s}} = 0.951^{+0.015}_{-0.019}$ (slope of the scalar perturbation
spectrum), $\tau = 0.09 \pm 0.03$ (optical depth), $\sigma_8 =
0.74^{+0.05}_{-0.06}$ (amplitude of fluctuations). 

Actually, it seems that there is no evidence for antimatter effects.  However,
the future Planck \citep{tauber03} mission will study anisotropies and
polarization structures of the CMB with unprecedented accuracy and the
antimatter signatures could be within its sensitivity.

\subsection{Neutrino experiments}

The exciting results of supernova neutrino measurements in 1987, that
opened the field of ``neutrino astronomy'' \citep{bahcall00}, were
obtained by experiments that have been built for other purposes (in
particular for the search of possible proton decays).  On the other
hand, the main issue of neutrino physics in the last dozen years is
bound to the neutrino oscillations: the measurements of the
differences between the squared masses of the mass eigenstates, that
are different from the flavor eigenstates, reached better and better
precisions \citep{altarelli03}.  

The next generation neutrino detectors\footnote{For a review of past,
  present and future neutrino detectors, see
  http://neutrinooscillation.org} can be grouped in two sets: the big
water/ice \v{C}erenkov detectors and the magnetic calorimeters.
The latter will be able to distinguish between the different leptons
produced by the interactions of neutrino fluxes emitted by particle
accelerators.  Hence, they will be able to measure the flavor of the
incident neutrinos, that can be compared to the known flavor of
neutrinos produced by the particle beams.  This is very important to
understand the details of neutrino oscillations, but it is also very
useful for neutrino astronomy.

In particular, supernova explosions, whose progenitor was an
antimatter star, differ significantly from those produced by matter
stars (see \S\ref{sec-neutrino}): the initial neutrino burst is due to
the antineutronization reaction, that produces
$\overline{\nu}_\text{e}$ instead of $\nu_\text{e}$.  Hence a flux of
neutrinos generated by the explosion of antimatter stars will produce
positrons through the reaction $\overline{\nu}_\text{e} + \text{p} \to
\particle{e^+} + \text{n}$, whereas neutrinos produced by matter
supernovae will create electrons ($\nu_\text{e} + \text{n} \to
\particle{e^-} + \text{p}$) in the detectors.  Inside a magnetic
field, electrons and positrons will be curved in opposite directions,
allowing for a good discrimination between the two cases.  The
drawback of these detectors is that, because the neutrino incident
beam has known energy and direction, their angular acceptance is quite
narrow compared to water \v{C}erenkov detectors, hence the supernova
detection probability is lower.  However, since the expected
interaction rate is quite low, these detectors might have the
possibility to be triggered also by off-beam events, in order to study
atmospheric and astrophysical neutrinos.

\subsection*{Acknowledgments}

The author wishes to thank Pavel Naselsky, Floyd W. Stecker and Maxim Khlopov
for their suggestions and corrections to the first draft of this work.

\bibliography{antimatter}

\end{document}